\documentclass[epj]{svjour}
\usepackage{graphicx}
\usepackage{amsmath}
\usepackage{booktabs}
\begin{document}
	\sloppy
	\title{Multifractal analysis of electronic states on random Voronoi-Delaunay lattices}
	\author{Martin Puschmann\inst{1}\and Philipp Cain\inst{1}\and Michael Schreiber\inst{1}\and Thomas Vojta\inst{2}}
	\authorrunning{Martin Puschmann \textit{et al.}}
	\institute{Institute of Physics, Technische Universit\"at Chemnitz, 09107 Chemnitz, Germany \and
		Department of Physics, Missouri University of Science and Technology, Rolla, Missouri 65409, USA}
	
	\date{Received:  / Revised version: }
	
	\abstract{We consider the transport of non-interacting electrons on two- and three-dimensional random Voronoi-Delaunay lattices. It was recently shown that these topologically disordered lattices feature strong disorder anticorrelations between the coordination numbers that qualitatively change the properties of continuous and first-order phase transitions. To determine whether or not these unusual features also influence Anderson localization, we study the electronic wave functions by multifractal analysis and finite-size scaling. We observe only localized states for all energies in the two-dimensional system. In three dimensions, we find two Anderson transitions between localized and extended states very close to the band edges. The critical exponent of the localization length is about 1.6. All these results agree with the usual orthogonal universality class. Additional generic energetic randomness introduced via random potentials does not lead to qualitative changes but allows us to obtain a phase diagram by varying the strength of these potentials.
	}
	\maketitle
	
	\section{Introduction}\label{sec:intro}
	
	More than 5 decades ago, Anderson \cite{And58} showed that random disorder can localize a quantum particle in space. This phenomenon, now called Anderson localization, and the corresponding Anderson transitions between localized and metallic phases have since attracted lots of experimental and theoretical attention (see, e.g., Refs.\ \cite{LeeR85,KraM93,EveM08} for reviews). Remarkably, important features of Anderson localization, such as the existence of a metallic phase and the qualitative characteristics of the Anderson transition, depend on the dimensionality and the symmetries of the system, in close analogy to conventional continuous phase transitions.
		
	This leads to a symmetry classification of Anderson transitions. Originally, three universality classes were identified, based on the invariance of the Hamiltonian under time reversal and spin rotations. These classes (orthogonal, unitary, and symplectic) correspond to the Wigner-Dyson classification of random matrices \cite{Wig51a,Dys62}. For example, the orthogonal universality class contains systems that are invariant under both time reversal and spin rotation. Later, this classification scheme was extended by including additional symmetries (for a review, see, e.g., Evers and Mirlin \cite{EveM08}). Within each class, the properties of the Anderson transition are expected to be universal. Specifically, all eigenstates in one-dimensional (1D) and two-dimensional (2D) systems in the orthogonal class are expected to be localized, implying the absence of an Anderson transition\footnote{Note that extended states (the so-called Azbel resonances \cite{Azb81,Azb83}) can be found at isolated values of the energy. As their measure is zero and as they appear randomly in the energy spectrum, they can usually be neglected.}. In contrast, the states in three-dimensional (3D) systems in the orthogonal class can undergo a transition from localized to delocalized as energy or disorder strength are varied; and this transition features universal critical exponents.
	
	These universal properties hold for uncorrelated randomness; spatial disorder correlations can lead to different behavior. Some short-range correlations have been shown to produce extended states at specific, isolated energies in an otherwise	localized system, for instance in the so-called dimer model \cite{DunWP90}. In contrast, long-range	correlations can lead to the appearance of a true metallic phase (and thus an Anderson transition), even in 1D. Some of these developments are reviewed by Izrailev et al. \cite{IzrKM12}.
	
	In recent years, topologically disordered systems (lattices with random connectivity) have attracted particular attention because phase transitions in such systems feature surprising violations	of the expected universal behavior. This includes transitions in Ising and Potts magnets as well as the contact process, all defined on random Voronoi-Delaunay (VD) lattices \cite{JanV95,LimCAA00,JanV02,LimCF08,OliAFD08}.	 Barghathi and Vojta \cite{BarV14} solved this puzzle by showing that the 2D random VD lattice belongs to a broad class of random lattices whose disorder fluctuations feature strong anticorrelations and	therefore decay faster with increasing length scale than those of generic random systems. Such lattices are ubiquitous in 2D because the Euler equation for a 2D graph imposes a topological	constraint on the coordination numbers; however, examples in higher dimensions exist as well. The suppressed disorder fluctuations lead to important modifications of the Harris \cite{Har74} and Imry-Ma \cite{ImrM75,ImrW79,HuiB89,AizW89} criteria that govern the effects of disorder on continuous and first-order transitions, respectively.
	
	These results immediately pose the question of whether or not the unusual features of random VD lattices also modify universal properties of Anderson localization. Grimm et al.\ \cite{GriRS98}	studied the energy level distribution of a tight-binding model defined on a random tessellation \cite{Cae91a,Cae91b} similar to a VD lattice. They found level repulsion indicative of extended states in the metallic regime. However, to the best of our knowledge, a systematic finite-size scaling (FSS) study that would permit the unambiguous identification of the metallic and localized phases	(and the Anderson transition between them) has not yet been performed.
	
	In this paper, we therefore investigate noninteracting electrons on 2D and 3D random VD lattices. We perform FSS based on a multifractal analysis (MFA) of the electronic wave functions. Our results can be summarized as follows. In 2D, we observe localized states for all energies. In contrast, the 3D system features two Anderson transitions between localized and extended states close to the band edges. The critical exponent of the correlation length takes the value $\nu \approx 1.6$, in agreement with the standard orthogonal universality class. This implies that the unusual coordination number anticorrelations of random VD lattices do not lead to qualitatively different behavior compared to the well-known Anderson model of localization on regular lattices.
	
	The rest of this paper is organized as follows. In Sec.\ \ref{sec:Anderson}, we introduce the random VD lattices and define the Hamiltonian. We also discuss FSS as well as the MFA of the electronic states. Section \ref{sec:Results} is devoted to the results of our simulations for both 2D and 3D systems. We conclude in Sec.\ \ref{sec:Conclusion}.
	
	\section{Model and methods}\label{sec:Anderson}
	\subsection{Random Voronoi-Delaunay lattices}
	
\begin{figure}
			\centering
			\includegraphics[]{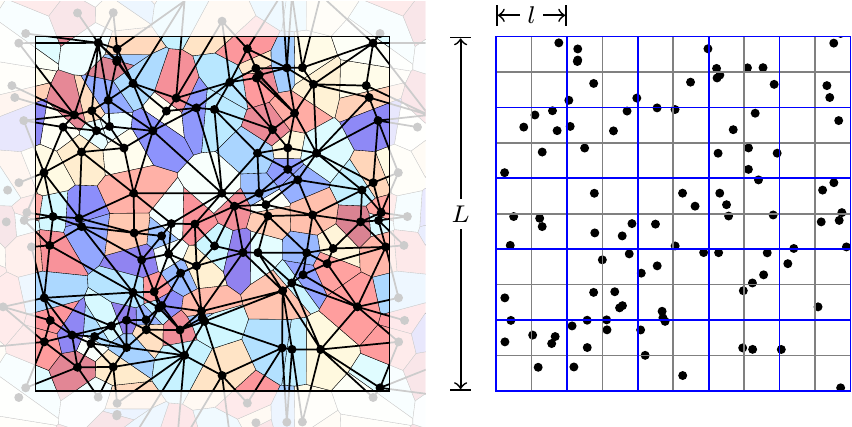}
			\caption{Left: 2D Voronoi-Delaunay construction for $N=100$ randomly positioned sites (dots) placed in a square of linear size $L=10$.
				The grayed outer margins show the periodic continuation of the original system (inner square). The Voronoi cells (polygons) are colored
				arbitrarily. The lines connecting lattice sites show the Delaunay triangulation.
				Right: MFA scheme for the same setup. The system is partitioned into boxes of linear size $l$. The smallest box size $l_\mathrm{min}=1$ is chosen such that each box contains one lattice site on average.}
			\label{fig:vdl_construct}
		\end{figure}

	The random VD lattice is a prototypical system with topological (connectivity) disorder. It can be viewed as a simple model for amorphous solids, foams or biological cell structures. The random VD lattice is defined as a set of lattice sites at random positions together with the bonds that connect nearest neighbor sites. These neighbors are determined by the VD construction \cite{OkaBSC00} as follows. The entire system area (or volume, in the 3D case) is subdivided into disjoint polygons (Voronoi cells) containing exactly one lattice site such that each cell contains all points in space that are closer to that lattice site than to any other site. The Voronoi diagram is the complete set of these Voronoi cells. Lattice sites whose Voronoi cells share an edge (or a face, in the 3D case) are considered neighbors, irrespective of their real-space distance. In 2D, the graph of all bonds connecting pairs of neighbors consists of triangles only. It is called the Delaunay triangulation. The corresponding construction in 3D leads to the Delaunay tetrahedralization, a lattice consisting of tetrahedra only. We emphasize that random VD lattices are not bipartite, i.e., they cannot be divided into disjoint sublattices A and B such that each bond connects an A site to a B site. A 2D example of a VD construction can be seen in Fig.~\ref{fig:vdl_construct}.

	In the following, we consider independent, uniformly distributed random lattice sites of density unity contained in a square or cubic box of linear size $L$. We employ periodic boundary conditions. More details of the algorithm that we use to perform the VD construction are given in the Appendix.
	
\begin{figure}
				\centering
				\includegraphics[]{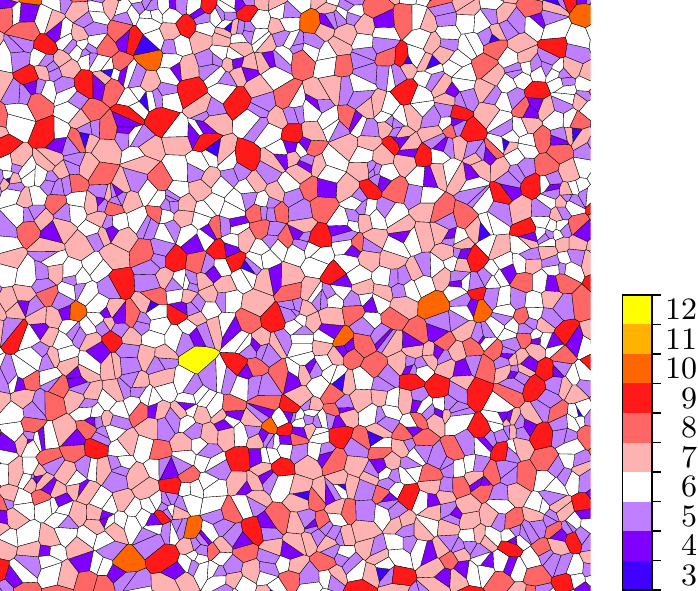}
				\caption{$40\times40$ section of a 2D random VD lattice. The color of each Voronoi cell represents the coordination number of the included lattice site.}
				\label{fig:2d_coordination}
			\end{figure}

\begin{figure}
				\includegraphics{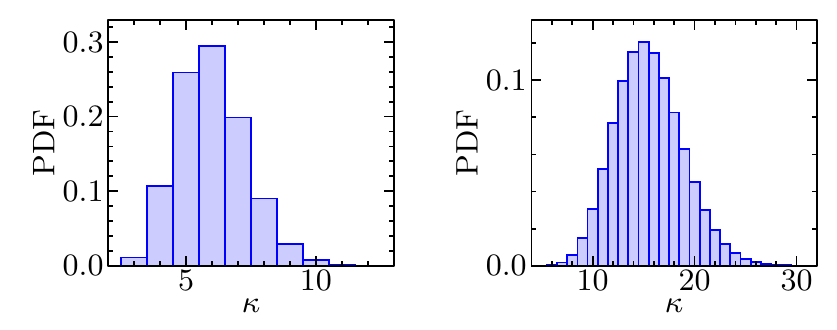}
				\caption{Probability distribution of the coordination numbers for 2D (left) and 3D (right) random VD lattices. The data are averages over $500$ systems of linear size $L=120$ for 2D and $L=24$ for 3D. Here, the maximal coordination numbers are $15$ and $35$ for the 2D and the 3D system, respectively. }
				\label{fig:coord_nmb}
			\end{figure}

	In a random VD lattice, the coordination number (number of neighbors) $\kappa_i$ fluctuates from site to site. This topological disorder is illustrated in Fig.~\ref{fig:2d_coordination}. The corresponding coordination number distribution can be seen in Fig.~\ref{fig:coord_nmb}. The average coordination number in 2D is exactly $\langle \kappa \rangle_\mathrm{2D}=6$. This is a consequence of the Euler equation for a 2D graph consisting of triangles only. In 3D, the average coordination number is given by $\langle \kappa \rangle_\mathrm{3D}=2 + (48/35)\pi^2 \approx 15.54$ \cite{Mei53}. In both dimensions, the mean coordination numbers are higher than those of the regular square and cubic lattices often used in numerical localization studies. The standard deviations of the coordinations numbers are $\sigma_\kappa \approx 1.33$ in 2D and $\sigma_\kappa \approx 3.36$ in 3D. This means that the disorder is moderately strong.
		
	\subsection{Anderson model}
    \label{subsec:Anderson_model}
	
	We now consider the motion of noninteracting electrons on a random VD lattice. We describe it by means of a tight-binding model having one (Wannier) orbital per lattice site. The corresponding Hamiltonian
		\begin{align}
		H=\sum\limits_{\langle i,j \rangle} \left|i\right\rangle \left\langle j\right|+ \sum\limits_{i} \upsilon_i\left|i\right\rangle \left\langle i\right| 	 \label{eq:Hamiltonian}
		\end{align}
	 is analogous to the Anderson model of localization. The first term describes the hopping of electrons between nearest (Voronoi) neighbors. The hopping matrix element is constant and fixed at unity. This term contains the topological disorder. The second term represents additional	 energetic randomness. The $\upsilon_i$ are random on-site potentials uniformly distributed in the interval $\left[-W/2,W/2\right]$. We are mostly interested in the case $W=0$ for which the disorder in the system is purely topological. However, for comparison with the usual Anderson model of localization, we also consider nonzero random potential strength $W$. 

To compute the densities of states (DOS), we directly diagonalize the secular matrices. The system sizes are $L=120$ for 2D and $L=24$ for 3D, limited by computer memory. To find the eigenstates of the Hamiltonian (\ref{eq:Hamiltonian}) close to a particular energy value for larger system sizes, we use a sparse matrix algorithm based on the Jacobi-Davidson method \cite{BolN06}. Here, we treat system sizes up to $L=2000$ for 2D and $L=140$ for 3D.
	
\begin{figure}
			\includegraphics{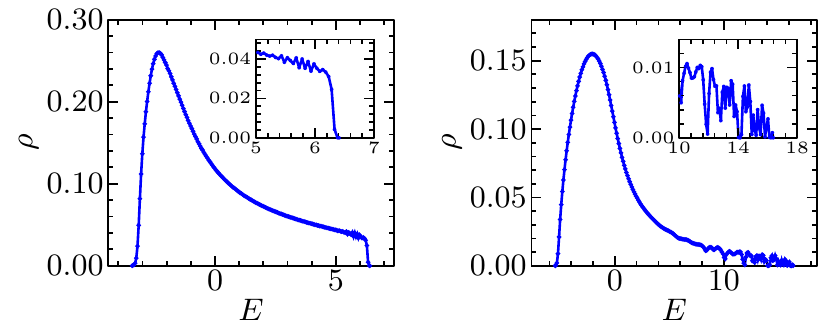}
			\caption{Density of states for the tight-binding Hamiltonian (\ref{eq:Hamiltonian}) on 2D (left) and 3D (right)	random VD lattices for the case of purely topological disorder, $W=0$. The data are averages over $500$ systems of linear size $L=120$ for 2D and $L=24$ for 3D. The insets show magnifications of the right band edges.}
			\label{fig:dos_W0}
		\end{figure}

	Figure \ref{fig:dos_W0} shows an overview over the DOS resulting from these calculations for purely topological disorder ($W=0$).
	As the VD lattices are not bipartite, the DOS is not symmetric with respect to the energy $E=0$. While the DOS near the low-energy band edge increases rapidly, there is a pronounced tail on the high-energy side, in particular in 3D. Interestingly, this tail does not stretch much beyond $E=\langle \kappa \rangle$. States in the far tail live on rare large clusters of sites with above average coordination numbers. The fact that there are almost no states with $E > \langle \kappa \rangle$ reflects the strong disorder anticorrelations of the random VD lattices that prevent the formation of large rare regions \cite{BarV14}. We also note that the DOS in the 3D band tail fluctuates significantly. We believe this is a finite-size effect.

\subsection{Finite-size scaling}
The scaling approach is a phenomenological description of the behavior close to the critical point of a continuous transition.
We consider a general dimensionless measure $\Lambda(E,L)$ characterizing the electronic states
as function of energy $E$ and a characteristic length, e.g., the system size $L$. Close to the
critical energy $E_\mathrm{c}$, it fulfills the scaling form \cite{SleO99a,OhtSK99}
		\begin{align}
		\Lambda(E,L)=F\left[\Omega_\mathrm{r}L^\frac{1}{\nu},\Omega_\mathrm{i}L^{-y}\right]~.
		\end{align}
The description contains the relevant scaling variable $\Omega_\mathrm{r}$ associated with the relevant exponent $\nu$ as well as the leading
irrelevant scaling variable $\Omega_\mathrm{i}$ associated with the irrelevant exponent $-y$. The dependence of the scaling
variables on $E$ can be described in terms of the expansions
		\begin{align}
		\Omega_\mathrm{r}&=\omega+\sum\limits_{n=2}^{m_\mathrm{r}}\frac{b_n}{n!}\omega^n\quad\text{and}\\
		\Omega_\mathrm{i}&=1+\sum\limits_{n=1}^{m_\mathrm{i}}\frac{c_n}{n!}\omega^n
		\end{align}
where $\omega={(E-E_\mathrm{c})}/{E_\mathrm{c}}$ is a dimensionless measure of the distance from the critical energy.
The scaling function
		\begin{align}		 F\left[\Omega_\mathrm{r}L^\frac{1}{\nu},\Omega_\mathrm{i}L^{-y}\right]=\sum\limits_{n=0}^{n_\mathrm{i}}\frac{\Omega_\mathrm{i}^nL^{-ny}}{n!}F_n\left[\Omega_\mathrm{r}L^\frac{1}{\nu}\right]
		\end{align}
	is expanded into a Taylor polynomial of the irrelevant variable. The coefficients are the regular single-variable scaling functions
		\begin{align}
		F_n\left[\Omega_\mathrm{r}L^\frac{1}{\nu}\right]&=\sum\limits_{k=0}^{n_\mathrm{r}}a_{n,k}T_k(\Omega_\mathrm{r}L^\frac{1}{\nu})\quad.
		\end{align}
	They are expanded into Chebyshev polynomials $T_k(\cdot)$ of the first kind and the $k$th order. This general expansion depends on the four expansion orders $n_\mathrm{r}$, $n_\mathrm{i}$, $m_\mathrm{r}$, and $m_\mathrm{i}$ yielding $(n_\mathrm{r}+1)(n_\mathrm{i}+1)+m_\mathrm{r}+m_\mathrm{i}+1+\Theta(n_\mathrm{i})$ free parameters. Their values are found by weighted nonlinear fits to the numerical data. The weights are formed by the reciprocal variances of the data points. This formalism allows us to simultaneously determine $E_\mathrm{c}$, $\nu$, and $y$.
	
	In order to get error estimates of these values we use a Monte-Carlo method \cite{RodVSR11}. It consists in building at least $10^4$ synthetic data sets by adding noise to the original values. This noise is created by Gaussian random numbers with a standard deviation $\sigma$ equal to the individual error of each data point.
By fitting $\Lambda$ to these synthetic data sets we obtain distributions for the parameter values. These distributions are usually Gaussian. Large deviations, e.g., superpositions of multiple Gaussians, are interpreted as instabilities. The corresponding regressions are not well defined and will be neglected. (This could be avoided by increasing the accuracy of the data, changing the number of parameters, or varying the initial conditions.) We note that this method constitutes an error-propagation calculation for the random errors only. Systematic errors are not detected completely, but the influence of different expansion orders can be identified.

\subsection{Multifractal analysis}\label{sec:MFA}
	Multifractal behavior is a feature of eigenstates at critical points~\cite{CasP86}. The MFA is based on a standard box-counting algorithm. The boxes are squares or cubes for 2D and 3D systems, respectively. The $d$-dimensional system of size $L^d$ is partitioned into boxes of size $l^d$ (see Fig.~\ref{fig:vdl_construct}). Using the probability
		\begin{align}
		\mu_b(\Phi,l,L)=\sum\limits_{r\in \mathrm{box}~b}\left|\Phi_r\right|^2\quad.
		\end{align}
	to find the electron in the $b$th box, the measure
		\begin{align}
		P_q(\Phi,l,L)=\sum\limits_{b}\mu^q_b(\Phi,l,L)
		\end{align}
	is constructed from its $q$th moment. Depending on $q$, this measure is dominated by boxes with either large or small $\mu_b$. For multifractal wave functions,
   $P_q$ is expected to behave as a power of  the normalized box size $\lambda=l/L$, with the scaling exponent (mass exponent)
		\begin{align}
		\tau_q(\Phi,l,L)=\lim\limits_{\lambda\rightarrow 0}\frac{\ln P_q(\Phi,l,L)}{\ln\lambda}\quad.\label{eqn:massexponent}
		\end{align}
	The singularity spectrum $f(\alpha_q)$ is the Legendre transform of $\tau_q$, and it comprises the scaling exponents of the fractal dimensions of all moments. A parametric representation of the singularity spectrum can be obtained by calculating the singularity strength
		\begin{align}
		\alpha_q=\frac{\mathrm{d}\tau_q}{\mathrm{d}q}=\lim\limits_{\lambda\rightarrow 0}\frac{\sum_{b}\bar{\mu}^q_b\ln\mu_b}{\ln\lambda}\label{eqn:singularity_strength}
		\end{align}
	and the fractal dimension
		\begin{align}
		f_q=q\alpha_q-\tau_q=\lim\limits_{\lambda\rightarrow 0}\frac{\sum_{b}\bar{\mu}^q_b\ln\bar{\mu}^q_b}{\ln\lambda}\label{eqn:fractal_dimension}
		\end{align}
	with
		\begin{align}
		\bar{\mu}_b^q=\frac{\mu_b^q}{P_q(\Phi,l,L)}\quad.
		\end{align}
	For statistical purposes, we utilize the ensemble average of these exponents~\cite{EveM08,VasRR08,RodVR08}. In particular, we use the ensemble averaged singularity strength
		\begin{align}
		\alpha_q=\lim\limits_{\lambda\rightarrow 0}\frac{A_q(l,L)}{\ln\lambda}\label{eqn:massexponent_ens}
		\end{align}
	with
		\begin{align}
		A_q(l,L)=\frac{\left\langle\sum_b \mu_b^q\ln\mu_b \right\rangle_\Phi}{\left\langle\sum_b \mu_b^q\right\rangle_\Phi}
		\end{align}
	to perform the finite-size scaling analysis outlined in the last section. $\langle\cdot\rangle_\Phi$ denotes the average over different eigenstates. When determining the error of $A_q$, one has to take into account that the numerator and denominator are correlated with each other~\cite{RodVSR11}.
	
Only integer box ratios $\lambda^{-1}=L/l$ are possible when partitioning the original system without overlap. A general ratio can be used by employing the periodic boundary conditions to fold back into the real system any protruding box parts  that arise for noninteger  $\lambda^{-1}=L/l$ \cite{SchG91,ThiS13}. To obtain a uniform sampling, we are then required to average over all possible box origins.
	
To extrapolate to the thermodynamic limit $\lambda\rightarrow 0$ in Eq. (\ref{eqn:massexponent_ens}), we perform a linear least-squares fit of the numerator versus the denominator at fixed $L$ (box-size scaling). The standard deviations $\sigma$ of $A_q(l)$ are utilized to determine the inaccuracy of the regression values. To approach the limit $\lambda\rightarrow 0$, sufficiently small values of $l$ have to be taken into account. However, for the smallest $l$, the data deviate from the power-law behavior because corrections to scaling
that stem from the irrelevant scaling variables become important. Their influence on the estimated exponent values can be reduced by optimizing the fit boundaries. We use $\lambda\in\left[0.1,0.3\right]$ for 2D and $\lambda\in\left[0.3,0.5\right]$ for 3D.
		\begin{figure}
			\includegraphics{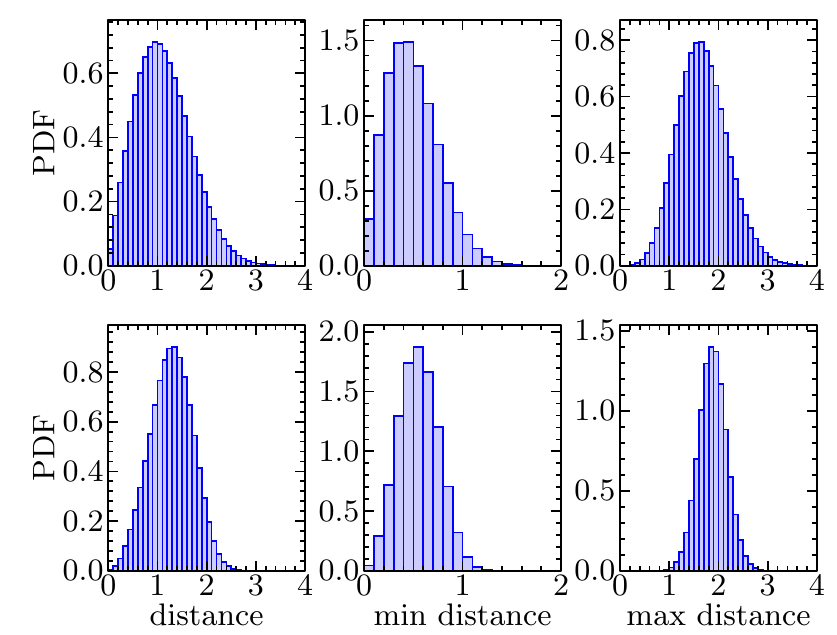}
			\caption{Probability distributions of the distances between neighboring sites: all (left), minimal (middle), and maximal (right) distances to neighboring sites. The data are averages over $500$ realizations of 2D (upper panels) and 3D (lower panels) VD lattices with $L=120$ and $L=24$, respectively.}
			\label{fig:mfa_dist}
		\end{figure}
			
In random VD lattices, the cells have varying sizes; the resolution of the MFA given by the smallest box size $l_\mathrm{min}$ is therefore arbitrary. We use $l_\mathrm{min}=1$, such that the average number of sites per box is $1$ (see Fig.~\ref{fig:vdl_construct}). Since empty boxes affect the MFA results, only box sizes larger than the maximum distance between neighboring sites are included in the MFA analysis. This guarantees that there are no empty boxes. In both dimensions, we use $l>4$ in accordance with the distributions shown in the right panel of Fig.~\ref{fig:mfa_dist}. 	
		\begin{figure*}
			\includegraphics{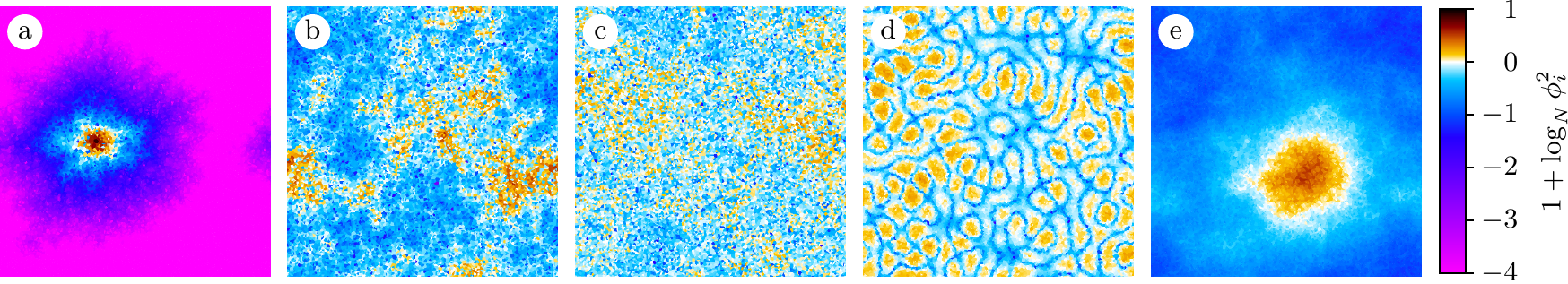}
			\caption{Eigenstates for different energies (from left to right: $\textrm{-}3.27$, $\textrm{-}2.89$, $0.00$, $5.80$, and $6.31$) of the same 2D VD lattice. Plots a) and e) show the wave functions of the lowest and highest eigenvalue, respectively. The amplitude $\phi^2_i$ of the individual site is given by the coloring of the Voronoi cells. The color white of the color scale corresponds to a homogeneously extended state, the color black corresponds to a state localized on a single site.}
			\label{fig:2d_eigstates}
		\end{figure*}

\section{Results}\label{sec:Results}
\subsection{2D system}
	Figure~\ref{fig:2d_eigstates} shows five representative eigenstates of the energy spectrum of a 2D system with $L=120$ and $W=0$. The wave function of the lowest eigenvalue
(Fig.\ \ref{fig:2d_eigstates}a) is exponentially localized. The maximum amplitude is concentrated at the single Voronoi cell with highest coordination number (here $\kappa=12$). States on the upper band edge also show localized behavior very clearly (Fig.\ \ref{fig:2d_eigstates}e). However, the highest amplitude is orders of magnitude smaller and, correspondingly, the localization length is larger. More generally, states of smaller energy are influenced by local fluctuations of the coordination number. States of higher energy are driven by interferences and the probability amplitudes behave as non-integrable Chladni figures~\cite{SteS92}. The number of anti\-nodes increases with decreasing eigenenergy. In summary, this is a visualization of localization in a classical and quantum mechanical manner for the states close to the lower and the upper band edge, respectively. Towards the band center the amplitude fluctuations increase and a classification by visual inspection is not easily possible anymore.

\begin{figure}
				\includegraphics{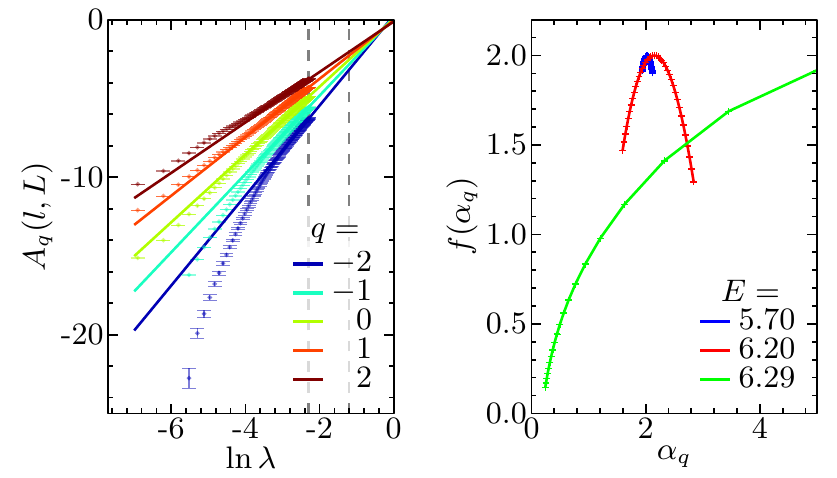}
				\caption{Left: Scaling of $A_q(l)$ with fixed $L=1000$ and different $q$ at $E=6.2$ and $W=0$. (Individual symbols are only shown for data points with $\lambda<0.1$.) All data with $\lambda\geq0.1$ follow the linear regressions (solid lines) closely. The dashed lines indicate the bounds of $\lambda\in \left[0,1,\ 0.3 \right]$ used for the fit. Right: Multifractal spectrum for different energies with moments $-2\leq q\leq 2$ staggered by $0.1$.}
				\label{fig:2D_alpha}
			\end{figure}

We therefore turn to the MFA to quantitatively characterize the eigenstates as function of energy.
Specifically, we consider the FSS behavior of $\alpha_q$ for systems from $L=60$ up to $2000$ with $1000$ states per data point. Figure\ \ref{fig:2D_alpha}
	shows the box-size scaling of $A_q(l,L)$ used to estimate $\alpha_q$ for several $q$. Whereas values at higher $\lambda$ show the expected power-law dependence on $\lambda$, results for small $l$ deviate systematically due to finite-size effects. Therefore, they are neglected as discussed before. The multifractal spectra $f(\alpha)$ resulting from the analysis are also shown in Fig.\ \ref{fig:2D_alpha} for representative energies close to the upper band edge. For the energy $E=6.2$, we obtain a parabola-like shape as expected. Towards the band center the parabola shrinks (see $E=5.7$) and approaches the limiting case for completely extended wave functions, namely the single point $(2,2)$. In the opposite direction, close to the band edge, the spectrum transforms towards the extremely localized limit, the points $(0,0)$ for negative $q$ and $(\infty,2)$ for positive $q$.

\begin{figure}
			\includegraphics{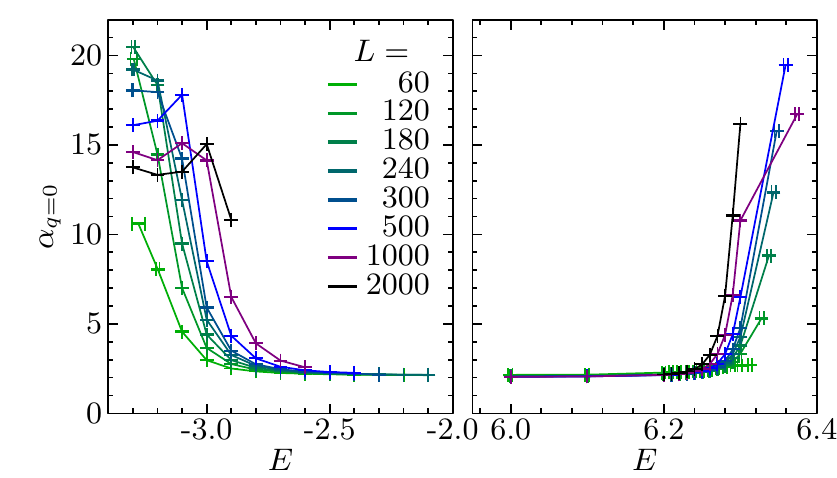}
			\caption{Scaling behavior of $\alpha_{q=0}(L,E)$ in 2D VD lattices for energies close to lower and upper band edge. Each data point is an average over $1000$ eigenstates close to the respective $E$. Error bars show $1\sigma$ intervals.}
			\label{fig:2D_alpha_ens_W_0}
		\end{figure}

The behavior of $\alpha_{q=0}(E,L)$ (the position of the maximum of the parabola) is shown in more detail in Fig.\ \ref{fig:2D_alpha_ens_W_0}.
	For energy values close to the band edge, $\alpha_{q=0}$ generally increases with $L$. This means that these states are localized. The strength of the localization depends on $E$. The pronounced localization at the lower band edge seen in Fig.\ \ref{fig:2d_eigstates} is reproduced here. Strongly fluctuating wave functions near the band center have a value close to $\alpha=2$ for all system sizes within their accuracy. The envelopes of corresponding wave functions have a very large localization length (larger than the system size). Importantly,
the curves  for different $L$ do not show a common crossing point, implying that there is no Anderson transition. All states are localized. (The seeming crossings at very large $\alpha_{q=0}(E,L)$ in Fig.\ \ref{fig:2D_alpha_ens_W_0} can be attributed to numerical artifacts.)
	
We have also studied the effects of additional potential disorder $W=2$ and $W=4$. It results in a broadening of the DOS and an enhancement of the localization.
However, qualitative changes compared to $W=0$ were not found. All states are localized.

\subsection{3D system}
\begin{figure}
			\includegraphics{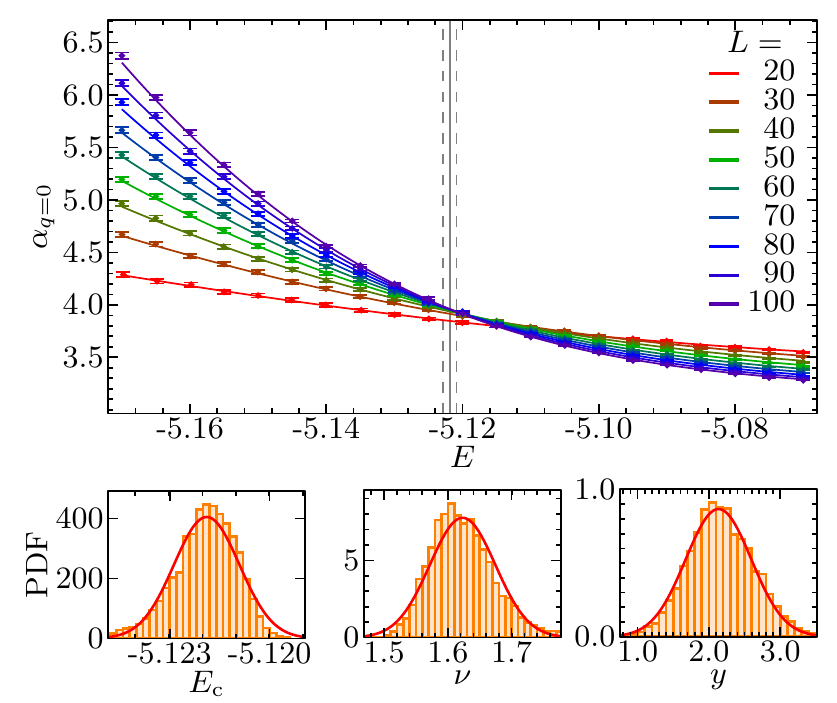}
			\caption{Scaling behavior of $\alpha_{q=0}$ in 3D VD lattices as function of system size $L$ and energy $E$ close to the lower band edge. Each data point is an average over $2000$ eigenstates of different lattices close to the respective $E$. Error bars show $1\sigma$ intervals. Lines are result of the scaling approach with $n_\mathrm{r}=4$, $n_\mathrm{i}=1$, $m_\mathrm{i}=0$, and $m_\mathrm{r}=3$ (see Tab.~1 for more details). The solid and dashed gray lines mark the estimated transition point and the corresponding $1\sigma$ error interval, respectively. Lower plots show the distributions of critical parameters obtained from  $10^4$ synthetic data sets. }
			\label{fig:3D_alpha_ens_W_0_ls}
		\end{figure}

\begin{figure}
			\includegraphics{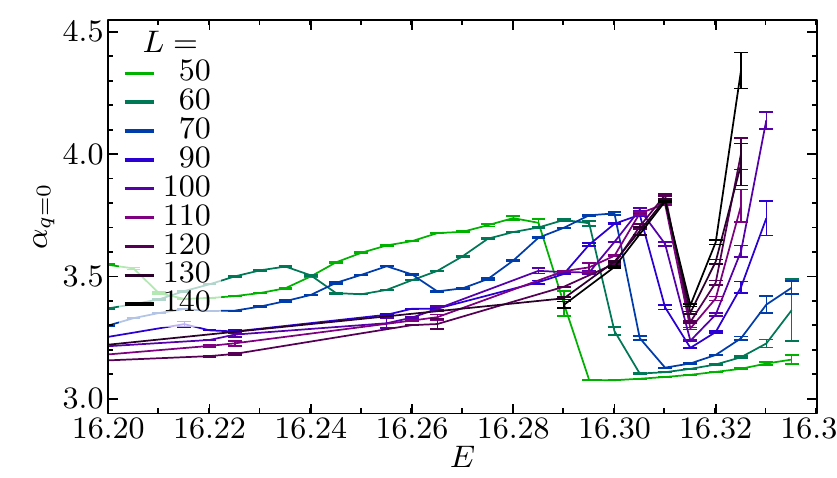}
			\caption{Behavior of $\alpha_q$ in 3D VD lattices as function of system size $L$ and energy $E$ close to the upper band edge. Number of considered eigenstates differs for different data points and is reflected in the error bars ($1\sigma$ intervals).}
			\label{fig:3D_alpha_ens_W_0_rs}
		\end{figure}

	The data analysis for the 3D VD lattice proceeds analogously. We use system sizes between $L=20$ and $L=140$. In contrast to the 2D system, we observe two Anderson transitions induced by purely topological disorder. These transitions are located  close to the two energy band edges. The corresponding FSS behavior of $\alpha_{q=0}$ is visualized in Fig.\ \ref{fig:3D_alpha_ens_W_0_ls} for the transition near the lower band edge and in Fig.\ \ref{fig:3D_alpha_ens_W_0_rs} for the upper edge. Thus, localized states exist only near the band edges. The broad central area of the energy band encompasses extended states. This corresponds to the fact that the topological disorder is only moderately strong and results in localization behavior similar to that of a weakly disordered regular Anderson model.

As shown in Fig.\ \ref{fig:3D_alpha_ens_W_0_ls}, the singularity strength $\alpha_q$ features a smooth energy dependence for the transition near the lower band edge. Thus, the FSS approach is applicable. The results are presented in Tab.~1. It can be seen that the critical parameters $E_\mathrm{c}, \nu$ and $y$ are influenced by the irrelevant scaling variable. We compare different expansion orders to demonstrate the stability of the regression results. In particular, we neglect the data of small systems $L<50$ and use a FSS approach without irregular expansion ($n_\mathrm{i}=0$). Such regressions show a higher robustness when changing initial conditions.
	Taking into account all data, we estimate the critical energy $E_\mathrm{c}=-5.122(3)$ and the critical exponent $\nu=1.62(4)$. The results of regressions without the irrelevant scaling variable deviate slightly ($E_\mathrm{c}=-5.121(1)$ and $\nu=1.58(4)$) because of the neglected systematic shift of the intersections described by the irrelevant term. The quality of all fits is very close to unity. This indicates that the errors of the original data points were overestimated.
		
	For the transition near the upper band edge, the $\alpha_{q=0}$ curves shown in Fig.\ \ref{fig:3D_alpha_ens_W_0_rs} are very noisy. This is caused by
the low DOS (see right panel of Fig.\ \ref{fig:dos_W0}) and finite size effects.
The existence of a transition can still be inferred because $\alpha_{q=0}$ decreases with $L$ for the smaller energies, while it
 increases with $L$ for the largest energies. This indicates a crossing and thus an Anderson transition. However, a clear transition point cannot be determined from the available data. In particular, the values are insufficient to perform a scaling analysis. (Also note that for small systems, gaps in the DOS appear. Therefore the number of considered eigenstates varies from $10$ to $8000$ between data points, leading to strong variations in the error bars.) A rough estimate of the critical energy is $E_\mathrm{c}=16.31$.

\begin{figure}
			\includegraphics{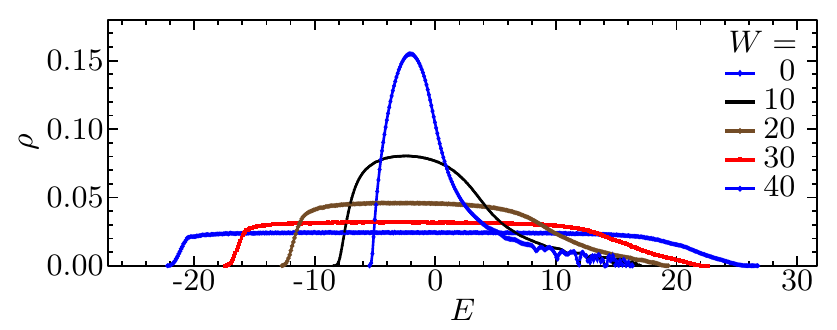}
			\caption{DOS for the 3D VD lattice with additional potential disorder. The data are averages over $500$ systems of linear size $L=24$.}
			\label{fig:3D_dos}
		\end{figure}

\begin{figure}
			\includegraphics{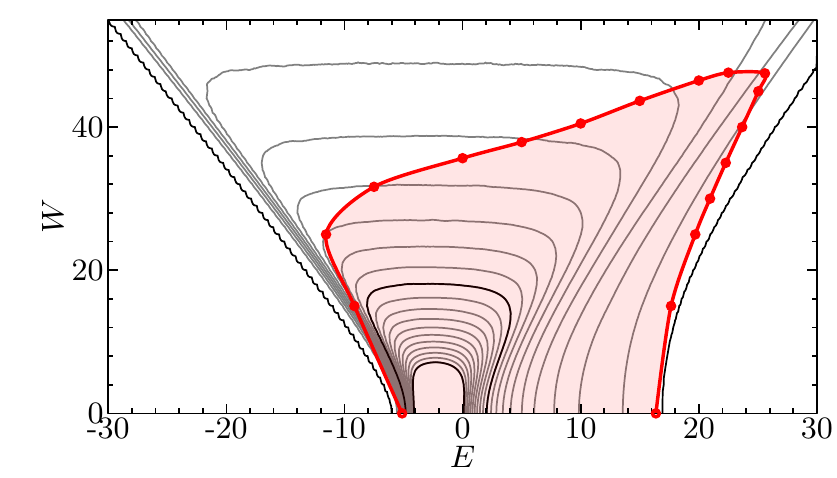}
			\caption{Phase diagram of the 3D VD lattice. Localized and delocalized (red shaded area) states are separated by the mobility edge (thick red line). The inaccuracy of the measured transitions (red dots) is smaller than the symbol size. The thin gray and black lines are contours of the DOS between $0.0$ and $0.1$ staggered by $0.005$.}
			\label{fig:phasediagram}
		\end{figure}

We now turn to the effects of additional on-site disorder, i.e., $W\ne 0$.
As in 2D, on-site disorder leads to a broadening of the DOS with increasing disorder strength $W$, this is demonstrated in Fig.\ \ref{fig:3D_dos}.
To find the localization phase diagram, we have performed a number of calculations with either fixed $E$ or fixed $W$.
System sizes $L=50,\ 60$, and $70$ were used with $10$ data points for each $L$ to determine the critical points.
Figure~\ref{fig:phasediagram} shows the resulting phase diagram.
The region of delocalized states is asymmetric. For positive energies, it extends towards higher disorder with a maximal disorder strength $W_\mathrm{c}^\mathrm{max}=47.5$ at $E=22$.
			\begin{figure}
				\includegraphics{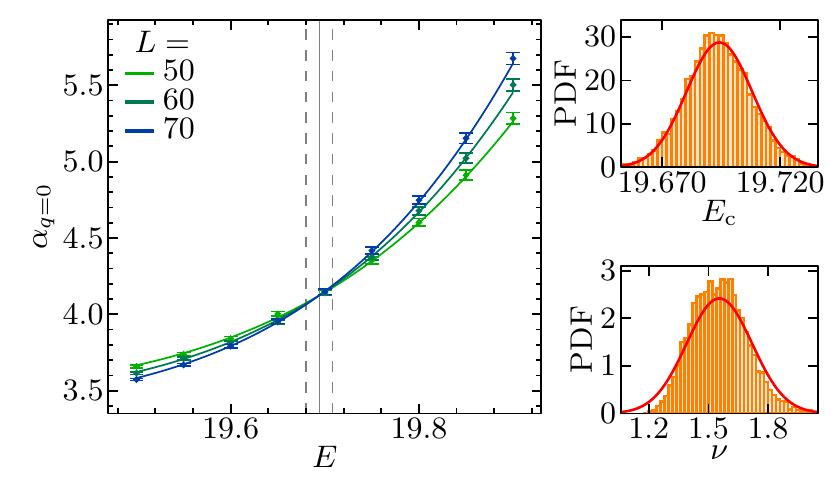}
				\caption{Scaling behavior (left) of $\alpha_{q=0}$ in 3D VD lattices with additional potential disorder $W=25$, as function of system size $L$ and energy $E$ close to the upper band edge. Each data point is an average over $1000$ eigenstates of different lattices close to the respective $E$. Error bars show $1\sigma$ intervals. Lines are result of the scaling approach with $n_\mathrm{r}=4$, $n_\mathrm{i}=0$, $m_\mathrm{i}=0$, and $m_\mathrm{r}=2$. The estimated exponent is $\nu=1.55(14)$. The solid and dashed gray lines mark the estimated transition point $E_\mathrm{c}=19.69(1)$ and the corresponding $1\sigma$ error interval, respectively. Right plots show the distributions of critical parameters obtained from $10^4$ synthetic data sets.}
				\label{fig:3D_alpha_ens_W_25_rs}
			\end{figure}
	Figure\ \ref{fig:3D_alpha_ens_W_25_rs} shows the localization transition near the upper band edge for fixed $W=25$. The additional potential disorder smoothes the density
of states and suppresses the artifacts in the $\alpha_{q=0}$ data seen for the pure 3D VD lattice. The FSS approach is thus applicable. We obtain $E_\mathrm{c}=19.69(1)$
and  $\nu=1.55(14)$.
	
	The transition induced by $W$ at fixed energy $E=10$ is studied in detail in order to obtain a more accurate estimate of $\nu$ for systems with both topological and potential disorder. The data analysis resembles the analysis of the transition tuned by $E$ for $W=0$, as described above. The details are summarized in Tab.~2. We observe a critical exponent $\nu=1.59(6)$ in agreement with the result for purely topological disorder.
			\begin{table*}
			\centering
			\label{tab:3D_W_0}
			\caption{Estimates of the critical parameters $E_\mathrm{c}$, $\nu$, and $y$ with their standard deviations for the transition near the lower energy band edge. The indices $\mathrm{l}$ and $\mathrm{u}$ denote the lower and upper bound of considered sizes, respectively. $Q$ is the quality of fit. $N_\mathrm{P}$ and $N_\mathrm{F}$ are the number of points and degrees of freedom, respectively. $\alpha_{q=0}$ was analyzed in the energy range $E\in[-5.17,\ -5.07]$ with $2000$ eigenstates per data point. We set $m_\mathrm{i}=0$ always. }			
			\begin {tabular}{ccccccccccccccc}%
		  	\toprule $E$&$\sigma _E$&$\nu $&$\sigma _\nu $&$y$&$\sigma _y$&$L_\mathrm {l}$&$L_\mathrm {u}$&$n_\mathrm {r}$&$n_\mathrm {i}$&$m_\mathrm {r}$&$\chi ^2$&$N_\mathrm {P}$&$N_\mathrm {F}$&$Q$\\\midrule %
		  	-5.1207 & 0.0005 & 1.580 & 0.0423 & -- & -- & 50 & 100 & 4 & 0 & 3 & 45.7 & 126 & 117 & 1.0000\\%
		  	-5.1206 & 0.0005 & 1.577 & 0.0404 & -- & -- & 50 & 100 & 4 & 0 & 4 & 45.5 & 126 & 116 & 1.0000\\%
		  	-5.1220 & 0.0010 & 1.626 & 0.0449 & 2.188 & 0.444 & 20 & 100 & 4 & 1 & 3 & 51.6 & 189 & 174 & 1.0000\\%
		  	-5.1220 & 0.0010 & 1.623 & 0.0494 & 2.157 & 0.447 & 20 & 100 & 4 & 1 & 4 & 51.6 & 189 & 173 &1.0000\\\bottomrule %
		  	\end {tabular}%
		  	\end{table*}
		\begin{table*}
			\centering
			\label{tab:3D_E_10}
			\caption{Same as Tab.~1, but for the transition at fixed $E=10$. $\alpha_{q=0}$ was analyzed in the disorder range $W\in[38.5,\ 43.5]$. }
			\begin {tabular}{ccccccccccccccc}%
				\toprule $W$ & $\sigma _W$ & $\nu $ & $\sigma _\nu $ & $y$ & $\sigma _y$ & $L_\mathrm {l}$ & $L_\mathrm {u}$ & $n_\mathrm {r}$ & $n_\mathrm {i}$ & $m_\mathrm {r}$ & $\chi ^2$ & $N_\mathrm {P}$ & $N_\mathrm {F}$ & $Q$\\\midrule %
				40.4181 & 0.0301 & 1.591 & 0.0294 & -- & -- & 50 & 100 & 2 & 0 & 2 & 43.0 & 150 & 144 & 1.0000\\%
				40.4224 & 0.0318 & 1.586 & 0.0315 & -- & -- & 50 & 100 & 2 & 0 & 3 & 42.8 & 150 & 143 & 1.0000\\%
				40.3287 & 0.0772 & 1.585 & 0.0554 & 1.871 & 0.539 & 20 & 100 & 2 & 1 & 2 & 51.7 & 225 & 215 & 1.0000\\%
				40.2947 & 0.0827 & 1.607 & 0.0592 & 1.671 & 0.581 & 20 & 100 & 2 & 1 & 3 & 51.2 & 225 & 214 & 1.0000\\\bottomrule %
			\end {tabular}%
		\end{table*}
\section{Conclusion}\label{sec:Conclusion}

To summarize, we have studied the effects of topological disorder on Anderson localization. To this end, we have investigated the wave functions of noninteracting electrons on random Voronoi-Delaunay lattices by multifractal analysis and finite-size scaling. In two dimensions, there is no Anderson transition as all states are localized, even in the absence of extra potential disorder. Adding random potentials further enhances the localization. In contrast, in three dimensions, the topological disorder of the Voronoi-Delaunay lattice induces two Anderson transitions close to the edges of the energy band, with localized states in the tails and extended states in the bulk of the band. If extra random potentials are added, the region of extended states first broadens with the broadening density of states, but then it shrinks and vanishes at some critical random potential strength.

All these qualitative features agree with those of the usual Anderson model of localization. This means that the anticorrelations of the topological disorder \cite{BarV14} do not affect the universal properties of Anderson localization. This also holds for the critical behavior of the localization transition in three dimensions. The correlation exponents found in the present paper, viz., $\nu=1.62(4)$ for purely topological disorder and $\nu=1.59(6)$ for combined topological and energetic disorder, agree within their errors with high-precision results for the usual Anderson model of localization \cite{RodVSR11,SleO14}.

Why is Anderson localization not (qualitatively) affected by the disorder anticorrelations of the Voronoi-Delaunay lattice even though other continuous and first-order phase transitions are qualitatively changed and violate the usual Harris and Imry-Ma criteria? In the systems in which these violations have been found \cite{JanV95,LimCAA00,JanV02,LimCF08,OliAFD08}, the coordination number directly determines the local distance from the transition point because the effective coupling strength is simply the sum over the effects of all neighbors. Anticorrelations of the coordination numbers thus generate anticorrelated random-mass (or random-$T_\mathrm{c}$) disorder. Anderson localization is more complex. In particular, quantum interference effects are crucial, at least away from the band edges, and these effects are not captured by the coordination number alone. Note, however, that the electronic states in the band tails, where the localization is mostly classical, do seem to be influenced by the disorder anticorrelations (see discussion at the end of Sec.\ \ref{subsec:Anderson_model}). Clearly, more work will be necessary to fully resolve the effects of topological disorder on Anderson localization.

This work was supported by the NSF under Grant Nos.\ DMR-1205803. We acknowledge useful discussions with H. Barghathi.

\appendix
\section*{Appendix: Algorithm for creating random Voronoi-Delaunay lattices}\label{sec:appendix}

\begin{figure}
			\centering
			\includegraphics{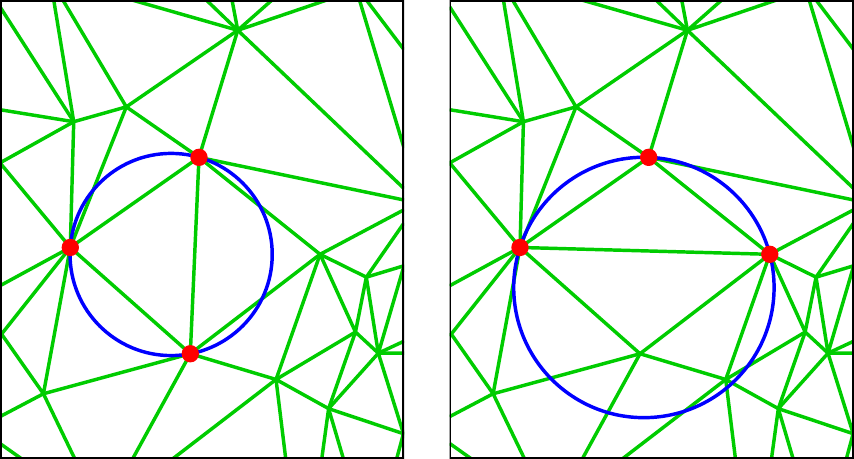}
			\caption{Left: Valid Delaunay triangulation, the circumcircle of the three chosen (red) sites does not contain	any other sites. Right: This triangulation of the same set of sites violates the empty circumcircle property because the circumcircle of the chosen triangle (formed by the red sites) does contain other sites.}
			\label{fig:circles2}
		\end{figure}

	Computing the Voronoi diagram or the Delaunay triangulation of a given set of points (lattice sites) is a standard problem of computational geometry, and many different algorithms are discussed in text books and the research literature (see, e.g., Ref.\ \cite{BerCKO08}). Our algorithm follows a suggestion by Tanemura et al. \cite{TanOO83} and is based on the remarkable ``empty circumcircle property'' of  a 2D Delaunay triangulation. It states that every triangle formed by the bonds (edges) of the Delaunay triangulation has an empty circumcircle, i.e., a circumcircle that does not contain any other lattice sites. This is illustrated in Fig.\ \ref{fig:circles2}.
	Analogously, a 3D Delaunay tessellation features an ``empty circumsphere property'': Each of the tetrahedra making up the tessellation has a circumsphere that does not contain any other lattice sites.
	
	Our algorithm considers the lattice sites one by one and finds the list of its (Voronoi) neighbors. This is done in a two-step process: (i) We first identify candidates for the neighbors based on their distance. All sites within a distance $R_\mathrm{c}$ from the given site are included in the candidate list. For optimal performance, the cutoff radius $R_\mathrm{c}$ should be chosen as small as possible without missing neighbors, reasonable values depend on the structure of the set of lattice sites under consideration (see below). (ii) From these candidates, we then construct all triangles (in 2D) with empty circumcirles for which the given site is one of the vertices. In 3D, we construct all tetrahedra  with empty circumspheres for which the given site is a vertex. The entire algorithm can be coded efficiently in less than 400 lines of Fortran 90 in 2D and less than 500 lines in 3D.
	
	For the current project, we have applied these algorithms to sets of $L^2$ (2D) or $L^3$ (3D) uniformly distributed random sites of density unity contained in a square or cubic box of linear size $L$. The sizes range up to $L=2000$ in 2D and $L=140$ in 3D, and we use periodic boundary conditions. We found that the necessary values of the cutoff radius $R_\mathrm{c}$ are quite small because the bond-length distribution of the random VD lattice drops off rapidly with increasing distance, see Fig.\ \ref{fig:mfa_dist} (in 2D, the tail is approximately Gaussian). Empirically, we found that $R_\mathrm{c}=5.7$ in 2D and $R_\mathrm{c}=3.7$ in 3D are sufficient to find all neighbors in all the lattices we considered.
	
	For system sizes $L > R_\mathrm{c}$, the computational effort of our algorithm scales approximately linearly with the number of sites. To give an example of the performance, finding the Delaunay triangulation of $10^6$ sites in 2D takes about 30 seconds on an Intel core i5-3570 CPU while $10^6$ sites in 3D take about 3 minutes.

\end{document}